\documentclass[conference]{IEEEtran}
\IEEEoverridecommandlockouts
\usepackage{cite}
\usepackage{amsmath,amssymb,amsfonts}
\usepackage{algorithmic}
\usepackage{graphicx}
\usepackage{textcomp}
\usepackage{xcolor}

\def\BibTeX{{\rm B\kern-.05em{\sc i\kern-.025em b}\kern-.08em T\kern-.1667em\lower.7ex\hbox{E}\kern-.125emX}}

\begin{document}

\title{Joint Beamforming for Intelligent Reflecting Surface-Assisted Millimeter
Wave Communications}

\author{Peilan Wang$^\dag$, Jun Fang$^\dag$, and Hongbin Li $^{\S}$,  \emph{Fellow, IEEE} \\
$^\dag$University of Electronic Science and Technology of China, Chengdu, China\\
$^{\S}$Stevens Institute of Technology, Hoboken, USA,\\
 \IEEEauthorblockA{Email: peilan.wangle@gmail.com, JunFang@uestc.edu.cn, and Hongbin.Li@stevens.edu}
}

\maketitle

\begin{abstract}
Millimeter wave (mmWave) communications are evolving as a
promising technology to meet the ever increasing data rate
requirements. However, high directivity and severe path loss make
it vulnerable to blockages, which could be frequent in indoor or
urban environments. To address this issue, intelligent reflecting
surfaces (IRSs) are introduced to provide additional adjustable
reflected paths. Most prior works assume that elements of IRSs
have an infinite phase resolution, which is difficult to be
realized in practical systems. In this paper, IRSs with
low-resolution phase shifters are considered. We aim to maximize
the receive signal power at the user by jointly optimizing
discrete phase shifts of IRSs and the transmit beamforming vector
at the base station for mmWave downlink systems. An analytical
near-optimal solution is developed by exploiting some important
characteristics of mmWave channels. Our theoretical analysis
reveals that low-resolution phase shifters can still achieve a
receive signal power that increases quadratically with the number
of reflecting elements. Simulation results are provided to
corroborate our analysis and show the effectiveness of the
proposed solution.
\end{abstract}

\begin{IEEEkeywords}
Millimeter wave communication, intelligent reflecting surfaces
(IRSs), beamforming.
\end{IEEEkeywords}

\section{Introduction}
MmWave communications are considered as a potential and promising
technology to support multi-gigabit wireless applications
\cite{RappaportMurdock11,GhoshThomas14}. However, mmWave signals
cannot diffract as well as their sub-$6$GHz counterparts and easily get blocked by
obstacles \cite{AbariBharadia17}, which is especially the case for indoor or dense urban environments \cite{TanSun18}. To address this issue,
intelligent reflecting surface (IRS) has recently emerged as a
promising and cost-effective solution to establish robust mmWave
connections even when the line-of-sight (LOS) link is blocked by
obstructions \cite{TanSun18,WangFang19}. IRS is a planar
array consisting of a large number of passive elements, each of
which can reflect the incident signal with a reconfigurable phase
shift and amplitude via a smart controller
\cite{CuiQi14,LiaskosNie18}. By smartly tuning the phase shifts of
passive elements, IRSs can help create effective virtual LOS
links, resulting in a more reliable mmWave connection
\cite{TanSun18,WangFang19}.

IRS-aided wireless communications have attracted much attention
recently \cite{BasarDi-Renzo19}. Prior works on IRS-assisted
transmissions can be found in
\cite{WuZhang18a,WuZhang18b,LiBin19,WangFang19,HuangZappone18,NadeemKammoun19}.
For the single-user scenario, it was shown
\cite{WuZhang18b,WangFang19} that IRSs are able to achieve a
squared power gain in terms of the number of reflecting elements,
thus creating a ``signal hotspot'' in the vicinity of the IRS. In
the multi-user scenario, \cite{HuangZappone18,WuZhang18a} claimed
that, by carefully adjusting the phase shift parameters of the
IRS, an ``interference-free'' zone can be formed near the IRS to
suppress interference for each user. Also, it was shown
\cite{NadeemKammoun19} that the IRS-assisted system can achieve
massive MIMO like gain with much fewer active antennas. However,
most of the above studies were based on the assumption that
elements of IRSs have an infinite phase resolution. Several works
considered IRS-aided systems with discrete phase shifts, the
proposed algorithms either involve an exhaustive search
\cite{TanSun18} or an iterative procedure to jointly search for
the optimal beamforming vector and discrete phase shift parameters
\cite{WuZhang19a,HuangGeorge18c}.

In this paper, we consider an IRS-assisted mmWave downlink system,
where multiple IRSs with discrete phase shifters are deployed to
assist the downlink transmission from a multi-antenna base station
to a single-antenna user. Our objective is to maximize the receive
signal power at the user by joint optimizing the phase shift
parameters of each IRS and the transmit beamforming vector at the
base station. Although such an optimization problem is generally
non-convex, by exploiting some inherent characteristics of mmWave
channels, we show that a near-optimal analytical solution can be
developed. Our theoretical analysis reveals that, even with
low-resolution phase shifters, our proposed solution can still
achieve a receive signal power that increases quadratically with
the number of reflecting elements. In addition, when compared with
the receive power achieved by IRSs with infinite-resolution phase
shifters, the receive signal power attained by our proposed
solution decreases by a constant factor that depends on the number
of quantization levels.


\section{System Model and Problem Formulation} \label{sec:system-model}
\begin{figure}[htbp]
\centerline {\includegraphics[width=6cm]{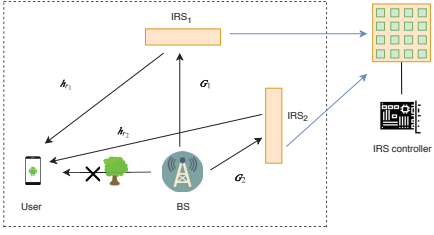}}
\vspace{-0.3cm} \caption{ Multiple-IRS assisted downlink MISO
system.} \label{fig1} \vspace{-0.4cm}
\end{figure}

As shown in Fig. \ref{fig1},  we consider an IRS-assited mmWave
downlink system, where multiple IRSs are deployed to assist the
communication from the base station (BS) to a single-antenna user.
Suppose $K$ IRSs are employed to assist the downlink transmission,
and each IRS is equipped with $M$ reflecting elements. The BS is
equipped with $N$ antennas. Denote $\boldsymbol{G}_k \in
\mathbb{C}^{M \times N}$ as the channel from the BS to the $k$th
IRS, and $\boldsymbol{h}_{r_k} \in \mathbb{C}^{M}$ as the channel
from the $k$th IRS to the user.

Each reflecting element of the IRS can reflect the incident signal
with a reconfigurable phase shift and amplitude via a smart
controller \cite{WuZhang19a}. Define
\begin{align}
\boldsymbol{\Theta}_k\triangleq\text{diag}(\beta_{k,1}
e^{j\theta_{k,1}},\ldots,\beta_{k,M} e^{j\theta_{k,M}})
\end{align}
as the phase-shift matrix of the $k$th IRS, where $\theta_{k,m}
\in [0,2\pi]$ and $\beta_{k,m} \in [0,1]$ denote the phase shift
and amplitude reflection coefficient associated with the $m$the
passive element of the $k$th IRS, respectively. For simplicity, we
assume $\beta_{k,m} = 1, \forall k, \forall m$ in the sequel of
this paper. Also, due to the hardware limitation, the phase shift
cannot take an arbitrary value, instead, it has to be chosen from
a finite set of discrete values \cite{WuZhang19a,TanSun18}.
Specifically, the set of discrete values for the phase shift is
defined as
\begin{align}
\theta_{k,m} \in {\mathcal{F}} \triangleq \left \{ 0, \frac{2\pi}{2^b} ,\ldots, \frac{2\pi (2^b-1)}{2^b} \right \}
\end{align}
where $b$ denotes the resolution of the phase shifter. In our
model, the direct link from the BS to the user is neglected due to
unfavorable propagation conditions. Denote $\boldsymbol{w} \in
\mathbb{C}^{N \times 1}$ as the beamforming/precoding vector adopt
by the BS. The signal received at the user is given by
\begin{align}
y= \left(\sum_{k=1}^K \boldsymbol{h}_{r_k}^H
\boldsymbol{\Theta}_k \boldsymbol{G}_k
\right)\boldsymbol{w} s + \epsilon \label{received-signal-model}
\end{align}
where $s$ is the transmitted signal modeled as a random variable
with zero mean and unit variance,  and $\epsilon$ denotes the
additive white Gaussian noise with zero mean and variance
$\sigma^2$. It should be noted that in the above model, we ignore
the signals reflected by two or more times due to the high path
attenuation of mmWave transmissions. Thus, the received signal
power at the user is given as
\begin{align}
\gamma = \left | \left(\sum_{k=1}^K \boldsymbol{h}_{r_k}^H
\boldsymbol{\Theta}_k \boldsymbol{G}_k
\right)\boldsymbol{w} \right |^2
\label{gamma}
\end{align}
Assuming the knowledge of global channel state information, we aim
to jointly design the transmit beamforming vector and the diagonal
phase-shift matrices $ \{\boldsymbol{\Theta}_k \}$ to maximize the
received signal power, i.e.
\begin{align}
\max_{\boldsymbol{w},\{\boldsymbol{\Theta}_k\}}\quad &
\left|\bigg(\sum_{k=1}^K \boldsymbol{h}_{r_k}^H
\boldsymbol{\Theta}_k \boldsymbol{G}_k
\bigg)\boldsymbol{w}\right|^2 \nonumber\\
\text{s.t.} \quad & \|\boldsymbol{w}\|_2^2\leq p \nonumber\\
& \boldsymbol{\Theta}_k=\text{diag}( e^{j\theta_{k,1}},\ldots,
e^{j\theta_{k,M}}) \quad \forall k \nonumber \\
&  \theta_{k,m} \in {\mathcal{F}}   \quad \forall k,\forall m
\label{opt1}
\end{align}
where $p$ denotes the maximum transmit signal power at the BS.

Note that such a joint beamforming problem has been studied in our
previous work \cite{WangFang19}, where the rank-one structure of BS-IRS channels
was exploited to help obtain a near-optimal analytical solution.
Nevertheless, \cite{WangFang19} assumes elements of IRSs have an infinite phase
resolution. Also, the direct link between the BS and the user was
considered in \cite{WangFang19}. In this work, following \cite{WangFang19}, we assume that the
channel from the BS to each IRS is a rank-one matrix, i.e.
\begin{align}
\boldsymbol{G}_k=\lambda_k\boldsymbol{a}_k\boldsymbol{b}_k^T \quad
\forall k \label{Gk}
\end{align}
where $\lambda_k$ is a scaling factor accounting for the complex
path gain and the antenna gain, $\boldsymbol{a}_k\in\mathbb{C}^{M
\times 1}$ and $\boldsymbol{b}_k\in\mathbb{C}^{N \times 1}$
represent the normalized array response vector associated with the
IRS and the BS, respectively. Such a rank-one channel assumption
is reasonable in mmWave communication systems because the power of
the mmWave LOS path is much higher (about 13dB higher) than the
sum of power of NLOS paths, as suggested by measurement campaigns
conducted in \cite{Muhi-EldeenIvrissimtzis10}. In practice, with
the knowledge of the location of the BS, IRSs can be installed
within sight of the BS.

On the other hand, due to the use of a large number of antennas at
the BS in mmWave systems and the fact that different IRSs, as seen
from the BS, are sufficiently separated in the angular domain, it
is reasonable to assume that array response vectors
$\{\boldsymbol{b}_k\}$ are near orthogonal to each other, i.e.
$|\boldsymbol{b}_i^H\boldsymbol{b}_j|\approx 0$ for $i\neq j$.


\section{Proposed solution} \label{sec:Multi-IRS}
In this section, we propose a near-optimal analytical solution for
the nonconvex problem (\ref{opt1}) by exploiting the rank-one
structure of BS-IRS channels and the near-orthogonality between
steering vectors $\{\boldsymbol{b}_k\}$. To solve (\ref{opt1}), we
first relax the discrete constraint placed on variables
$\{\theta_{k,m}\}$:
\begin{align}
\max_{\boldsymbol{w},\{\boldsymbol{\Theta}_k\}}\quad &
\left|\bigg(\sum_k^K \boldsymbol{h}_{r_k}^H \boldsymbol{\Theta}_k
\boldsymbol{G}_k
\bigg)\boldsymbol{w}\right|^2 \nonumber\\
\text{s.t.} \quad & \|\boldsymbol{w}\|_2^2\leq p \nonumber\\
& \boldsymbol{\Theta}_k=\text{diag}( e^{j\theta_{k,1}},\ldots,
e^{j\theta_{k,M}}) \quad \forall k
\label{opt2}
\end{align}
Substituting
$\boldsymbol{G}_k=\lambda_k\boldsymbol{a}_k\boldsymbol{b}_k^T$
into the objective function of (\ref{opt1}), we arrive at
\begin{align}
\quad &\left|\bigg(\sum_{k=1}^K \boldsymbol{h}_{r_k}^H
\boldsymbol{\Theta}_k \boldsymbol{G}_k
\bigg)\boldsymbol{w}\right|^2 = \left|\bigg(\sum_{k=1}^K
\lambda_k\boldsymbol{h}_{r_k}^H \boldsymbol{\Theta}_k
\boldsymbol{a}_k\boldsymbol{b}_k^T  \bigg)\boldsymbol{w}\right|^2
\nonumber
\\
\stackrel{(a)}{=}& \left|
\sum_{k=1}^K\eta_k\boldsymbol{\theta}_k^T\boldsymbol{g}_k\right|^2
\stackrel{(b)}{=}\left|
\sum_{k=1}^K\eta_k\boldsymbol{\bar{\theta}}_k^T\boldsymbol{g}_k
e^{j\alpha_k}\right|^2
\nonumber\\
\stackrel{(c)}{\leq}& \sum_{k=1}^K
\left|\eta_k\boldsymbol{\bar{\theta}}_k^T\boldsymbol{g}_k\right|^2
+\sum_{i=1}^K \sum_{j\neq i}^K
|\eta_i\boldsymbol{\bar{\theta}}_i^T
\boldsymbol{g}_i|\cdot|\eta_j\boldsymbol{\bar{\theta}}_j^T\boldsymbol{g}_j|
\label{upper-bound}
\end{align}
where in $(a)$, we define
$\eta_k\triangleq\boldsymbol{b}_k^T\boldsymbol{w}$,
$\boldsymbol{g}_k\triangleq
\lambda_k(\boldsymbol{h}_{r_k}^{\ast}\circ\boldsymbol{a}_k)$,
$\circ$ denotes the Hadamard (elementwise) product, and
$\boldsymbol{\theta}_k\triangleq
[e^{j\theta_{k,1}}\phantom{0}\ldots\phantom{0}
e^{j\theta_{k,M}}]^T$, in $(b)$, we express
$\boldsymbol{\theta}_k=\boldsymbol{\bar{\theta}}_k e^{j\alpha_k}$,
and the inequality $(c)$ becomes equality when arguments (i.e.
phases) of all complex numbers are identical. It should be noted
that we can always find a set of $\{\alpha_k\}$ such that the
arguments of $\eta_k\boldsymbol{\bar{\theta_k}}^T\boldsymbol{g}_k
e^{j\alpha_k},\forall k$ are identical, although at this point we
do not know the values of $\{\alpha_k\}$. Therefore \eqref{opt2}
can be converted into the following optimization
\begin{align}
\max_{\boldsymbol{w},\{\boldsymbol{\bar{\theta}}_k\}}\quad &
\sum_{k=1}^K
\left|\eta_k\boldsymbol{\bar{\theta}}_k^T\boldsymbol{g}_k\right|^2+  \sum_{i=1}^K
\sum_{j\neq i}^K
|\eta_i\boldsymbol{\bar{\theta}}_i^T\boldsymbol{g}_i|\cdot|\eta_j\boldsymbol{\bar{\theta}}_j^T\boldsymbol{g}_j|
\nonumber\\
\text{s.t.} \quad & \|\boldsymbol{w}\|_2^2\leq p \label{opt-ub}
\end{align}
From (\ref{opt-ub}), it is clear that the optimization of
$\{\boldsymbol{\bar{\theta}}_k\}$ can be decomposed into a number
of independent sub-problems, with $\boldsymbol{\bar{\theta}}_k$
solved by
\begin{align}
\max_{\boldsymbol{\bar{\theta}}_k}\quad &
|\boldsymbol{\bar{\theta}}_k^T\boldsymbol{g}_k| \nonumber\\
\text{s.t.} \quad &
\boldsymbol{\bar{\theta}}_k=[e^{j\bar{\theta}_{k,1}}\phantom{0}\ldots\phantom{0}
e^{j\bar{\theta}_{k,M}}]^T
\end{align}
It can be easily verified that the objective function reaches its
maximum $\| \boldsymbol{g}_k\|_{1}$ when
\begin{align}
\boldsymbol{\bar{\theta}}_k^{\star} = [ e^{-j {\rm arg} (g_{k,1})}
\phantom{0}\ldots\phantom{0} e^{-j {\rm arg} (g_{k,M})} ]
\label{opt-theta-c}
\end{align}
where $g_{k,m}$ denotes the $m$th entry of $\boldsymbol{g}_k$, and
$\arg(x)$ denotes the argument of the complex number $x$.

So far we have obtained the optimal solution of
$\{\boldsymbol{\bar{\theta}}_k\}$, which is independent of
$\{\alpha_k\}$ and $\boldsymbol{w}$. Based on this result,
(\ref{opt2}) is simplified as optimizing $\boldsymbol{w}$ and
$\{\alpha_k\}$:
\begin{align}
\max_{\boldsymbol{w},\{\alpha_k\}}\quad & \left|\bigg(\sum_{k=1}^K
\lambda_k e^{j\alpha_k}\boldsymbol{h}_{r_k}^H
\boldsymbol{\bar{\Theta}}_k^{\star}
\boldsymbol{a}_k\boldsymbol{b}_k^T
\bigg)\boldsymbol{w}\right|^2 \nonumber\\
\text{s.t.} \quad & \|\boldsymbol{w}\|_2^2\leq p \label{opt7}
\end{align}
where
$\boldsymbol{\bar{\Theta}}_k^{\star}\triangleq\text{diag}(\boldsymbol{\bar{\theta}}_k^{\star})$.
Note that $\lambda_k\boldsymbol{h}_{r_k}^H
\boldsymbol{\bar{\Theta}}_k^{\star}\boldsymbol{a}_k =
\boldsymbol{g}_k^T
\boldsymbol{\bar{\theta}}_k^{\star}=\|\boldsymbol{g}_k\|_1\triangleq
z_k$ is a real-valued number. Thus the objective function of
(\ref{opt7}) can be written in a more compact form as
\begin{align}
\left|\bigg(\sum_{k=1}^K z_k e^{j\alpha_k} \boldsymbol{b}_{k}^T\bigg)\boldsymbol{w}\right|^2
\stackrel{(a)}{=}&\left| \boldsymbol{v}^H
\boldsymbol{D}_z\boldsymbol{B}   \boldsymbol{w}\right|^2
\stackrel{(b)}{=}  \left| \boldsymbol{v}^H \boldsymbol{
\Phi} \boldsymbol{w}\right|^2
\end{align}
where in $(a)$, we define $\boldsymbol{v}\triangleq
[e^{j\alpha_1}\phantom{0}\ldots\phantom{0} e^{j\alpha_K}]^H$,
$\boldsymbol{D}_z\triangleq \text{diag}(z_1,\ldots, z_K)$ and
$\boldsymbol{B}\triangleq[\boldsymbol{b}_1\phantom{0}\ldots\phantom{0}
\boldsymbol{b}_K]^T $, and in $(b)$, we define
$\boldsymbol{\Phi}\triangleq\boldsymbol{D}_z\boldsymbol{B}$. Hence
(\ref{opt7}) can be simplified as
\begin{align}
\max_{\boldsymbol{w},{\boldsymbol{v}}}\quad &
\left| \boldsymbol{v}^H  \boldsymbol{ \Phi}  \boldsymbol{w}\right|^2 \nonumber\\
\text{s.t.} \quad & \|\boldsymbol{w}\|_2^2\leq p
\nonumber\\
& \boldsymbol{v}= [e^{j\alpha_1}\phantom{0}\ldots\phantom{0}
e^{j\alpha_K}]^H \label{opt8}
\end{align}
Note that for any given $\boldsymbol{v}$, the optimal precoding
vector $\boldsymbol{w}$ is the maximum-ratio transmission (MRT)
solution, i.e. $\boldsymbol{w}^{\star} = \sqrt{p}{
\left(\boldsymbol{v}^H \boldsymbol{\Phi}\right)^H } /{\|
\boldsymbol{v}^H\boldsymbol{\Phi}\|_2}$. Substituting the optimal
precoding vector $\boldsymbol{w}^{\star}$ into (\ref{opt8}) yields
\begin{align}
 \max_{\boldsymbol{v}} \quad & \boldsymbol{v}^H \boldsymbol{\Phi \Phi}^H
 \boldsymbol{v} \nonumber \\
\text{s.t.} \quad & |v_k| = 1 \quad \forall k \label{opt9}
\end{align}
The above optimization is a non-convex quadratically constrained
quadratic program (QCQP). It was shown this QCQP can be relaxed as
a standard convex semidefinite program (SDP) \cite{SoZhang07}.
Nevertheless, such an approach is computationally expensive and
does not admit a closed-form solution. On the other hand, note
that $\{\boldsymbol{b}_k\}$ are near orthogonal to each other.
Hence we have
\begin{align}
\boldsymbol v^H \boldsymbol{\Phi \Phi}^H
\boldsymbol v  = \boldsymbol v^H \boldsymbol{D}_z
\boldsymbol{B}\boldsymbol{B}^H\boldsymbol{D}_z\boldsymbol{v}  \approx \| \boldsymbol z\|_2^2 \label{eqn8}
\end{align}
which is a constant independent of the vector $\boldsymbol{v}$.
Hence any vector $\boldsymbol{v}$ which satisfies the constraint
$|v_k|=1,\forall k$ is a near-optimal solution to (\ref{opt9}).
For simplicity, we choose $\alpha_k = 0,\forall k$ as a solution
to (\ref{opt9}). In this case, we have
\begin{align}
\boldsymbol{\theta}_k^{\star}=\boldsymbol{\bar{\theta}}_k^{\star}
\quad \forall k
\end{align}
where $\boldsymbol{\bar{\theta}}_k^{\star}$ is given by
(\ref{opt-theta-c}).

Considering the finite resolution constraint imposed on the phase
shifters, each phase shift, $\theta_{k,m}$, can take on a discrete
value that is closest to its optimal value $\theta_{k,m}^{\star}$:
\begin{align}
\theta_{k,m}^{\ast} =  \arg\min_{\theta \in \mathcal{F}} \quad
|\theta-\theta_{k,m}^{\star}| \label{opt-theta}
\end{align}
where $\theta_{k,m}^{\star}$ denotes the $m$th entry of
$\boldsymbol{\theta}_k^{\star}$. After the phase shifts are
determined, the beamforming vector $\boldsymbol{w}$ can be
obtained according to the MRT solution.

\section{Performance Analysis} \label{sec-analysis-1}
In this section, we analyze the power scaling law of the average
received power as $M \rightarrow \infty$. For simplicity, we set
the maximum transmit signal power $p=1$. From (\ref{gamma}), the
average received power attained by our solution with $b$-bit phase
shifters is given by
\begin{align}
\gamma(b) &= \mathbb E \bigg[ \bigg \| \sum_{k=1}^K
\boldsymbol{h}_{r_k}^H \boldsymbol{\Theta}_k^{\ast}
\boldsymbol{G}_k \bigg\|_2^2 \bigg] \label{gammab}
\end{align}
where
$\boldsymbol{\Theta}_k^{\ast}=\text{diag}(\theta_{k,1}^{\ast},\ldots,\theta_{k,M}^{\ast})$
with $\theta_{k,m}^{\ast}$ given by (\ref{opt-theta}). Our main
results are summarized as follows.
\newtheorem{proposition}{Proposition}
\begin{proposition}
Assume $\boldsymbol {h}_{r_k}\sim {\cal CN}(0, \varrho_{r_k}^2
I)$, and the BS-$k$th IRS channel is characterized by (\ref{Gk}),
with $\lambda_k=\sqrt{N M}\rho_k$, where $\rho_{k}$ denotes the
complex path gain. As $M \rightarrow \infty$, we have
\begin{align}
\eta(b) \triangleq \frac{\gamma(b)}{\gamma(\infty)} =
\left(  \frac{2^b}{\pi} \sin\left( \frac{\pi}{2^b}\right)\right)^2
\label{eta}
\end{align} \label{proposition1}
\end{proposition}
\begin{IEEEproof}
Substituting (\ref{Gk}) into (\ref{gammab}), we arrive at
\begin{align}
\gamma(b)   & = \mathbb E \bigg[ \bigg\| \sum_{k=1}^K
\sqrt{NM}\rho_k\boldsymbol{h}_{r_k}^H
\boldsymbol{\Theta}_k^{\ast}\boldsymbol{a}_k\boldsymbol{b}_k^T \bigg\|_2^2 \bigg]          \nonumber \\
&\stackrel{(a)}=  \mathbb E \bigg[ \bigg \| \sum_{k=1}^K \tilde{z}_k  \boldsymbol{b}_k^T    \bigg\|_2^2  \bigg] \nonumber \\
&\stackrel{(b)} \approx  \sum_{k=1}^K \mathbb E[\tilde{z}_k^2]
\end{align}
where $(b)$ comes from the fact that $\{\boldsymbol{b}_k\}$ are
near orthogonal to each other, and in $(a)$, we define
\begin{align}
\tilde{z}_k \triangleq \sqrt{NM}\rho_k\boldsymbol{h}_{r_k}^H
\boldsymbol{\Theta}_k^{\ast}\boldsymbol{a}_k =
\sqrt{N}|\rho_k|\cdot \sum_{m=1}^M |h_{r_{k,m}}|e^{j\Delta
\theta_{k,m}}
\end{align}
in which $h_{r_{k,m}}$ is the $m$th entry of the channel vector
$\boldsymbol{h}_{r_k}$, and $\Delta \theta_{k,m}$ is the
discretization error
\begin{align}
\Delta \theta_{k,m} \triangleq {\theta}_{k,m}^{\ast}-
\theta_{k,m}^{\star}
\end{align}
Since discrete phase shifts in $\mathcal{F}$ are uniformly spaced,
discretization errors $\{\Delta \theta_{k,m} \}$ can be considered
as independent random variables uniformly distributed on the
interval $[ -\pi / 2^b, \pi / 2^b]$. Therefore we have
\begin{align}
\mathbb E [\tilde{z}_k^2] = &  N \mathbb E[|\rho_k|^2]  \mathbb
E\bigg[ \sum_{m=1}^M
|h_{r_{k,m}}|^2 \bigg] \nonumber \\
&  +N \mathbb E[|\rho_k|^2] \mathbb E \bigg[\sum_{m=1}^M
\sum_{i \neq m}^M |h_{r_{k,m}}| |h_{r_{k,i}}|  e^{j(\Delta \theta_{k,m}- \Delta \theta_{k,i})} \bigg] \nonumber \\
\end{align}
in which
\begin{align}
&\mathbb E \bigg[\sum_{m=1}^M |h_{r_{k,m}}|^2 \bigg] = M \varrho_{r_k}^2 \\
& \mathbb E [ e^{j\Delta \theta_{k,m}} ] = \mathbb E [ -e^{j\Delta \theta_{k,m} }] =
\frac{2^b}{\pi} \sin{\left( \frac{\pi}{2^b} \right)}
\end{align}
Finally, the average received power can be given as
\begin{align}
\gamma(b) \approx &
  NM \sum_{k=1}^K  \varrho_{r_k}^2 \mathbb E[|\rho_k|^2]  \nonumber \\
&  +  N M(M-1) \sum_{k=1}^K \mathbb E[|\rho_k|^2]  \frac{{\pi}
\varrho_{r_k}^2} {4} \left(\frac{2^b}{\pi} \sin{\left( \frac{\pi}{2^b} \right)} \right)^2
\label{gamma-b}
\end{align}
It is not difficult to verify that $\mathbb E [e^{j\Delta
\theta_{k,m}}]$ increases monotonically with $b$ and approaches
$1$ as $b \rightarrow \infty$. Hence, it can be easily obtained
that the ratio of $\gamma(b)$ to $\gamma(\infty)$ is given by
(\ref{eta}) as $M \rightarrow \infty$. This completes the proof.
\end{IEEEproof}

This proposition provides a quantitative analysis of the average
received signal power in multiple-IRS assisted system with
discrete phases shifts. We see that, when compared with the
receive power achieved by IRSs with infinite-resolution phase
shifters, the receive signal power attained by our proposed
solution decreases by a constant factor that depends on the number
of quantization levels $b$. Specifically, we have $\eta(1) =
0.4053$, $\eta(2) = 0.8106$ and $\eta(3) = 0.9496$. This result
also implies that the squared improvement \cite{WangFang19}
brought by IRSs with continuous phase shifts can still be achieved
with low resolution phase shifters.  Note that in \cite{WuZhang19a}, the author obtained a similar insight for a single IRS assisted MISO system by assuming that the BS is equipped with only one antenna. We extend this conclusion to the multiple-IRS assisted MISO system based on the assumption of rank-one structure of the channel matrix between the BS and each IRS.

\begin{figure}[!t]
\centering {\includegraphics[width=6cm]{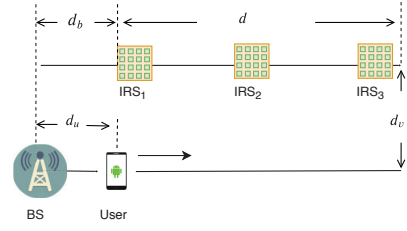}} \caption{
Simulation setup for multiple-IRS assisted system.}
\label{figsetup}
\end{figure}

\section{Simulation Results} \label{sec:experiments}
In this section, we present simulation results to illustrate the
performance of our proposed IRS-assisted beamforming solution. We
consider a scenario where the BS employs a uniform linear array
(ULA) with $N$ antennas, and each IRS consists of a uniform
rectangular array (URA) with $M = M_y M_z$ reflecting elements, in
which $M_y$ and $M_z$ denote the number of elements along the
horizontal axis and vertical axis, respectively. In our
simulations, the IRS-user channel is generated according to the
following geometric channel model \cite{AyachRajagopal14}:
\begin{align}
    \boldsymbol{h} = \sqrt{ \frac{M}{L} }\sum_{l=1}^{L}\alpha_{l}
\lambda_r\lambda_t\boldsymbol{a}_{t}(\phi_{a_l}, \phi_{e_l})
\label{channel-model}
\end{align}
where $L$ is the number of paths, $\alpha_{l}$ is the complex gain
associated with the $l$th path, $\phi_{a_l}$($\phi_{e_l}$) is the
associated azimuth (elevation) angle of departure,
$\boldsymbol{a}_{t}\in\mathbb{C}^{M}$ is the normalized transmit
array response vector, $\lambda_r$ and $\lambda_t$ denote the
receive and transmit antenna element gain. According to
\cite{WuZhang18a,FaisalNabil18}, $\lambda_r$ and $\lambda_t$ are
respectively set to $0$dBi and $9.82$dBi for the IRS-user link.
The complex gain $\alpha_l$ is generated according to a complex
Gaussian distribution \cite{AkdenizLiu14}
\begin{align}
\alpha_l\sim {\cal CN}(0,10^{-0.1\kappa}) \label{eqn9}
\end{align}
with $\kappa$ given as
\begin{align}
\kappa= e+10f \log_{10}(\tilde{d}) + \xi
\label{kappa}
\end{align}
in which $\tilde{d}$ denotes the distance between the transmitter
and the receiver, and $\xi\sim \mathcal{N}(0,\sigma_{\xi}^2)$. The
values of $e$, $f$, $\sigma_{\xi}$ are set to be $e=72$, $f=2.92$,
and $\sigma_{\xi}=8.7$dB, as suggested by real-world channel
measurements in the NLOS scenario at $28$GHz  \cite{AkdenizLiu14}. The
BS-IRS channel is characterized by a rank-one geometric channel
model given as
\begin{align}
\boldsymbol{G} = {\sqrt{NM}}\alpha\lambda_r\lambda_t \boldsymbol
{a}_r(\vartheta_{a},\vartheta_{e}) \boldsymbol{a}_t^H(\phi)
\end{align}
where $\vartheta_{a}$ ($\vartheta_{e}$) denotes the azimuth
(elevation) angle of arrival associated with the BS-IRS path,
$\phi$ is the associated angle of departure, $\boldsymbol{a}_r \in
\mathbb C^{M}$ and $\boldsymbol{a}_t \in \mathbb C^{N}$ represent
the normalized receive and transmit array response vectors,
respectively. In our simulations, $\lambda_r$ and $\lambda_t$ are
set to $0$dBi and $9.82$dBi, respectively. The complex gain
$\alpha$ is generated according to (\ref{eqn9}) and the
values of $e$, $f$, $\sigma_{\xi}$  in \eqref{kappa} are set to be $e=61.4$, $f=2$,
and $\sigma_{\xi}=5.8$dB, as suggested by real-world channel
measurements in the LOS scenario at $28$GHz  \cite{AkdenizLiu14}. Also, unless
specified otherwise, we assume $N=32$, $M_y=10$, and $M_z=5$ in
our experiments. Other parameters are set as follows: $p=30$dBm,
$\sigma^2=-85$dBm.

We consider a setup as depicted in Fig. \ref{figsetup}, where $K$
IRSs are equally spaced on a straight line which is in parallel
with the line connecting the BS and the user. Specifically, the
horizontal distance $d_b$ between the BS and the first IRS is set
to $d_b=11$m and the vertical distance is set to $d_v=1.5$m. Also,
the distance between the nearest IRS and the farthest IRS is set
to be $d=50$m.

To verify the near-optimality of our proposed solution, an upper
bound of the receive SNR can be obtained by solving a relaxed
convex formulation of (\ref{opt9}) \cite{WangFang19}. Note that
this upper bound is attained by assuming infinite-precision phase
shifters. Also, to show the benefits brought by IRSs, a
conventional system without IRSs is considered, where the channel
between the BS and the user is generated according to \cite{AyachRajagopal14}
\begin{align}
    \boldsymbol{h}_d = \sqrt{ \frac{N}{L} }\sum_{l=1}^{L}\alpha_{l}
\boldsymbol{a}_{t}(\phi_{l})
\label{channel-model}
\end{align}
where $L$ is the number of paths, $\alpha_{l}$ is the complex gain
associated with the $l$th path, $\phi_{l}$ is the
associated azimuth angle of departure,
$\boldsymbol{a}_{t}\in\mathbb{C}^{N}$ is the normalized transmit
array response vector. The complex gain $\alpha_q$ is generated according to \eqref{eqn9} and \eqref{kappa} in the NLOS scenario at $28$GHz  \cite{AkdenizLiu14}. 
The optimal MRT solution is employed to
achieve the maximum receive SNR.

\begin{figure}[!t]
\centering {\includegraphics[width=6cm]{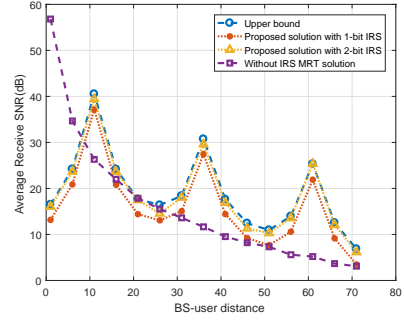}} \vspace{-0.3cm}
\caption{ Average receive SNR versus BS-IRS horizontal distance,
$K=3$.} \label{figSNRvDK3} \vspace{-0.4cm}
\end{figure}

\begin{figure}[!t]
\centering {\includegraphics[width=6cm]{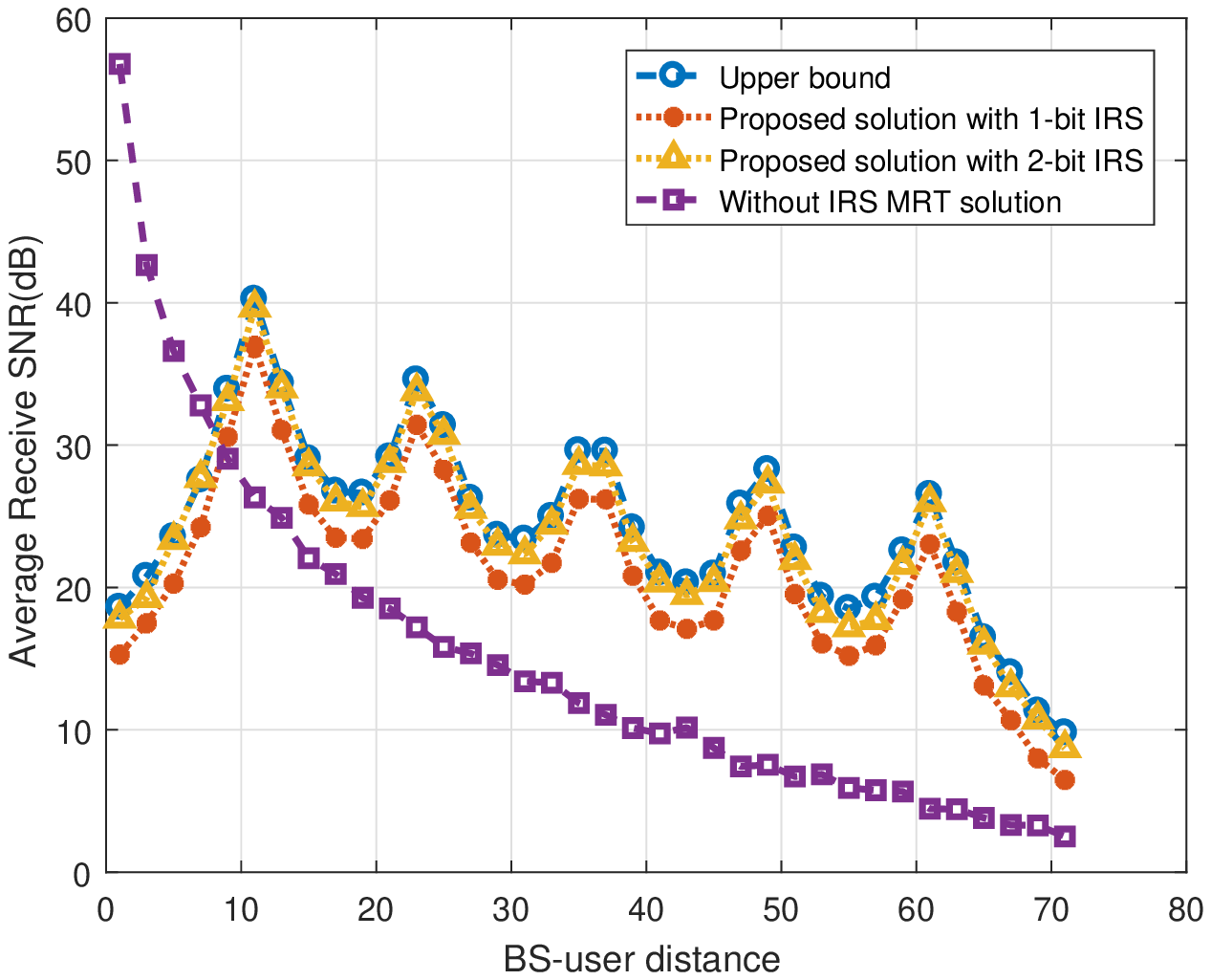}} \vspace{-0.3cm}
\caption{ Average receive SNR versus BS-IRS horizontal distance,
$K=5$.} \label{figSNRvDK5} \vspace{-0.4cm}
\end{figure}

Fig. \ref{figSNRvDK3} and Fig. \ref{figSNRvDK5} plot the receive
SNRs of our proposed solution and the MRT solution vs. the
distance between the BS and the user. It can be observed that the
IRS-assisted system can help substantially improve the receive
SNR, especially when the user is far away from the BS. Also, each
IRS creates a ``signal hotspot'' in its vicinity: when the user
moves closer to the IRS, its receive SNR becomes higher. In
addition, we see that with $2$-bit quantized phase shifts, our
proposed solution can achieve a receive SNR close to the upper
bound attained by assuming infinite-precision phase shifters. This
result validates the near-optimality of our proposed solution.

\begin{figure}[!t]
\centering {\includegraphics[width=6cm]{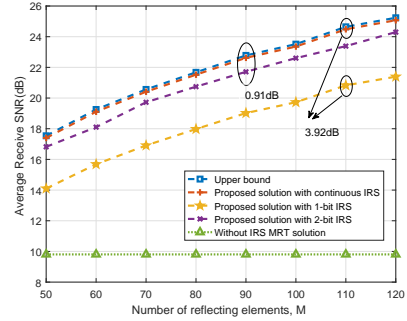}} \vspace{-0.3cm}
\caption{ Average receive SNR versus IRS reflecting elements.}
\label{figN2law} \vspace{-0.4cm}
\end{figure}

To verify our power scaling law analysis, we plot the receive SNRs
of different schemes versus the number of reflecting elements at
each IRS in Fig. \ref{figN2law}, where we set $K=3$, $d_u=41$m,
and we fix $M_y =10$ and change $M_z$. It can be observed that
even with discrete phase shifters, the squared improvement still
holds for our proposed solution. For example, when one-bit phase
shifters are used, the receive SNR at the user is about $14$dB
when $M=50$, and it gains $6$dB increase as $M$ doubles. Also, the
receive SNR loss due to the use of low-resolution phase shifters
is analyzed and given by (\ref{eta}). Specifically, we have
$\eta(1) = -3.9224$dB and $\eta(2)= -0.9121$dB. It can be observed
that simulation results are consistent with our theoretical
results.


\begin{figure}[!t]
\centering {\includegraphics[width=6cm]{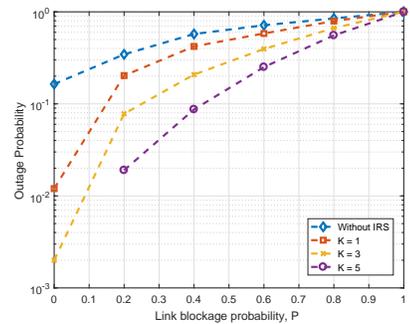}} \vspace{-0.3cm}
\caption{ Outage probability versus Link blockage probability.}
\label{figOutP} \vspace{-0.4cm}
\end{figure}

To show the robustness of IRS-assisted systems against blockages,
we calculate the outage probability as follows
\begin{align}
\mathbb{P}_{\text{out}}(\tau)= {\mathbb P} \left(\mathbb E \bigg[ 10 \log_{10}
\left(1+\frac{\gamma(b)}{\sigma^2} \right) \bigg]<\tau  \right)
\end{align}
where $\tau$ denotes the required threshold level and set to
$\tau=1.5$dB. For simplicity, we assume the link between BS and
each IRS is always connected and the link between each IRS and the
user is blocked with a pre-specified probability $P$. We set $d_u = 61$m and $b=2$. From Fig.
\ref{figOutP}, we observe that the outage probability can be
substantially reduced by deploying IRSs. Also, the more the IRSs
are deployed, the lower the outage probability can be achieved.


\section{Conclusions} \label{sec:conclusions}
In this paper, we studied the problem of joint active and passive
beamforming design for IRS-assisted mmWave systems, where multiple IRSs with discrete phase shifters are deployed to assist the downlink transmission from the BS to a single-antenna user. The objective is to maximize the received signal power by jointly optimizing the transmit beamforming vector and the discrete phase shift parameters at each IRS. By exploiting some inherent characteristics of mmWave channels, we derived a near-optimal analytical solution. Theoretical analysis reveals that low-resolution phase shifters can still achieve a receive signal power that increases quadratically with the number of reflecting elements at each IRS.  Simulations were provided to corroborate our analysis and illustrate the near-optimality of our proposed solution.

\section*{Acknowledgment}

This work was supported in part by the National Science
Foundation of China under Grant 61829103 and Grant 61871091.

\bibliography{newbib}
\bibliographystyle{IEEEtran}

\end{document}